\documentclass[aoas,MSNbibl,nameyear,seceqn,dvips]{arximspdf}
\usepackage{dcolumn,multirow}
\usepackage{graphicx}

\doi{10.1214/12-AOAS586} %kopijuoti is PTS
\volume{7}
\issue{1}
\pubyear{2013}
\firstpage{369}
\lastpage{390}

\makeatletter
\newcolumntype{d}[1]{D{.}{.}{#1}}
\def\bs{\mathbf}
\def\bss{\bolds}
\def\til{\widetilde}
\makeatother

\begin{document}
\begin{frontmatter}

\title{Efficient computation with a linear mixed model on large-scale
data sets with applications to genetic studies\thanksref{T1}}
\thankstext{T1}{Supported by the Wellcome Trust,
as part of the Wellcome Trust Case Control Consortium 2 project
[085475/B/08/Z and 085475/Z/08/Z]
and through the Wellcome Trust core Grant for the Wellcome Trust
Centre for Human Genetics [090532/Z/09/Z].
The use of the British 1958 Birth Cohort DNA collection was
funded by the Medical Research Council Grant G0000934
and the Wellcome Trust Grant 068545/Z/02, and of
the UK National Blood Service controls funded by the Wellcome Trust.}
\runtitle{Linear mixed models with large data sets}

\begin{aug}
\author[A]{\fnms{Matti} \snm{Pirinen}\corref{}\ead[label=e1]{matti.pirinen@iki.fi}},
\author[B]{\fnms{Peter} \snm{Donnelly}\ead[label=e2]{peter.donnelly@well.ox.ac.uk}\thanksref{t2}}
\and
\author[A]{\fnms{Chris C.~A.} \snm{Spencer}\ead[label=e3]{chris.spencer@well.ox.ac.uk}\thanksref{t3}}
\thankstext{t2}{Supported in part by a Wolfson Royal Society Merit
Award and
a Wellcome Trust Senior Investigator Award [095552/Z/11/Z].}
\thankstext{t3}{Supported in part by a Wellcome Trust Career
Development Fellowship [097364/Z/11/Z].}
\runauthor{M. Pirinen, P. Donnelly and C.~C.~A. Spencer}
\affiliation{University of Oxford}
\address[A]{M. Pirinen\\
C.~C.~A. Spencer\\
Wellcome Trust Centre for Human Genetics\\
University of Oxford\\
Roosevelt Drive\\
OX3 7BN\\
Oxford\\
United Kingdom\\
\printead{e1}\\
\hphantom{E-mail: }\printead*{e3}}
\address[B]{P. Donnelly\\
Department of Statistics\\
University of Oxford\\
1 South Parks Road\\
OX1 3TG\\
Oxford\\
United Kingdom\\
\printead{e2}}

\end{aug}

% HISTORY:
\received{\smonth{6} \syear{2011}}
\revised{\smonth{7} \syear{2012}}

% ABSTRACT
%
\begin{abstract}
Motivated by genome-wide association studies,
we consider a standard linear model with one additional random effect
in situations where many predictors have been collected on the same subjects
and each predictor is analyzed separately.
Three novel contributions are
(1) a transformation between the linear
and log-odds scales which is accurate for the important genetic case of
small effect sizes;
(2) a likelihood-maximization algorithm that is an order of
magnitude faster than the previously published approaches; and
(3) efficient methods for computing marginal likelihoods which allow
Bayesian model comparison.
The methodology has been successfully applied to a large-scale
association study of multiple sclerosis
including over 20,000 individuals and 500,000 genetic variants.
\end{abstract}

% KEYWORDS
%

\end{frontmatter}

%s1 #&#
\section{Introduction}\label{sec1}
We describe computationally efficient methods to analyze
one of the simplest linear mixed models:
%
%e1.1 #&#
\begin{equation}
\label{eqgenerativemodel} \bs Y=\bs X \bss\beta+\bss\varrho+\bss
\varepsilon,
\end{equation}
where $\bs Y=(y_1,\ldots,y_n)^T$ is the vector of responses on $n$ subjects,
$\bs X=(x_{ik})$ is the $n \times K$ matrix of predictor values on the subjects,
$\bss\beta=(\beta_1,\ldots,\beta_K)^T$
collects the (unknown) linear effects of the predictors on the
responses $\bs Y$
and the random effects $\bss\varrho$ and $\bss\varepsilon$ are
assigned the distributions
%
%e1.2 #&#
\begin{equation}
\label{eqpriorsvar} \bss\varrho|\bigl(\eta, \sigma^2
\bigr) \sim\mathcal{N}\bigl(0,\eta\sigma^2 \bs R\bigr) \quad\mbox{and}\quad
\bss\varepsilon|\bigl(\eta, \sigma^2\bigr) \sim\mathcal{N}
\bigl(0,(1-\eta)\sigma^2 \bs I\bigr).
\end{equation}
Here $\bs R$ is a known positive semi-definite $n\times n$ matrix,
$\bs I$ is the $n\times n$ identity matrix
and parameters $\sigma^2>0$ and $\eta\in[0,1]$
determine how the variance is divided between
$\bss\varrho$ and $\bss\varepsilon$.

Originally this model arose to
explain how the genetic component of a quantitative trait, such as height,
is correlated between relatives [\citet{Fisher1918}].
Many extensions of the model
have been thoroughly studied in genetics to estimate heritabilities
of traits, breeding values of individuals and locations of quantitative
trait loci
[see, e.g., \citet{Lynch1998,Sorensen2002}].

Recently, the model has been applied to
genome-wide association studies (GWAS)
[\citet{Astle2009}, Kang et al. (\citeyear{Kang2008,Kang2010}), \citet{Yu2005}, Zhang et al. (\citeyear{Zhang2009,Zhang2010})].
GWAS measure genotypes
at a large number (500,000--1,000,000) of
single-nucleotide polymorphisms (SNPs)
in large samples of individuals,
with the goal of identifying
genetic variants that explain
variation in a phenotype [\citet{McCarthy2008}].
Typically, GWAS data are analyzed
by testing each SNP separately using
standard linear or logistic regression models.
However, these models become invalid if the ascertainment procedure
itself introduces correlations between the phenotype and
the genetic background of the individuals.
[See \citet{Astle2009} for a detailed description of
spurious associations in GWAS.]
The linear mixed model (\ref{eqgenerativemodel})
can reduce the confounding effects
by using the covariance matrix $\bs R$ of the random effect $\bss
\varrho$
to model the genome-wide relatedness between the samples.
To emphasize the structure of the GWAS application,
we write the model as
%
%e1.3 #&#
\begin{equation}
\bs Y= \bs C \bss\beta_C +\bs X^{(\ell)} \bss
\beta_\ell+ \bss\varrho+\bss\varepsilon,
\end{equation}
where $\bs X^{(\ell)}$ contains the genetic data at the SNP $\ell$
and the matrix $\bs C$ contains the nongenetic covariates, such as age
and sex.
The most common strategy is to set $\bs X^{(\ell)}$ equal to the
number of copies of
the minor allele at the SNP $\ell$, but also dominant, recessive or
more complex
genetic effects can be modeled in this framework.
Even when the model needs to be analyzed for millions of different
$\bs X^{(\ell)}$ matrices, one for each SNP,
efficient computation becomes possible since the matrix $\bs R$
remains constant for a large number of the SNPs.

Our work with this model is motivated by a large GWAS on multiple sclerosis
(20,119 individuals, 520,000 SNPs), which we explain in detail
in Section~\ref{sec2}.
This case--control data set required novel methodological and computational
contributions which,
together with their applications in other genetics problems,
are explained in the remaining sections of this paper.

Section~\ref{sec3} gives a justification for applying
the linear mixed model to binary data and introduces a way to transform
the effect size estimates from the linear to log-odds scale.
Such a transformation is crucial for a meaningful interpretation of the
effect sizes and for combining the results with other separately
analyzed data sets,
for example, in a replication phase of GWAS or in a meta-analysis of
several independent studies.

The large size of the typical GWAS puts a premium on computational efficiency.
Section~\ref{sec4} describes a novel algorithm for likelihood analysis that reduces
the computation time from hundreds of years, as would be required by
the existing EMMA algorithm [\citet{Kang2008}], to only a few days and
is almost as fast as previous approximations
to the model [\citet{Kang2010,Zhang2010}].
With our implementation it is computationally feasible
to determine when the full model is noticeably more powerful than the
existing approximations as we demonstrate in Section~\ref{sec4}.

Bayesian approaches
provide a natural way to utilize
prior knowledge on the genetic architecture of
common diseases [\citet{Stephens2009}].
In Section~\ref{sec5} we compute Bayes factors using the linear mixed
model. The first application is in evaluating the genetic associations
in the multiple sclerosis data set.
The second application
investigates
when a nonzero heritability can be convincingly detected
in a large and only distantly related
population sample of individuals.

In the GWAS setting the challenge of combining data across genetically
heterogeneous
collections with strongly differing case--control ratios will
become more routine as study sizes increase. We therefore hope that our results
will be important in human genetics, and potentially also in
other fields of science, where large amounts of heterogeneous data need
to be analyzed
efficiently.

We have implemented the CM algorithm, the GLS approximation,
the log-odds estimation procedure and the Bayes factor computation
in software package MMM (\url{http://www.iki.fi/mpirinen}). The C-source code is publicly available under
the GNU General Public License.

%s2 #&#
\section{Motivating data set: Multiple sclerosis}\label{sec2}
Multiple sclerosis (MS) is a disease of the central nervous system
that can manifest itself through a variety of neurological symptoms
including, for example, motor problems, changes in sensation and
chronic pain.
The largest individual genetic effect is
associated with a region of the major histocompatibility complex on
chromosome 6,
and about 20 additional risk loci for MS had been identified by
the beginning of 2011.

Recently we were involved in a large GWAS of MS
[\citet{Sawcer2011}].
The study was divided into the UK component (1854 cases and 5175 controls)
and the non-UK component (7918 cases and 12,201 controls) which were
analyzed separately and combined via a fixed-effects meta-analysis.
About 100 of the most promising signals
among the 470,000 SNPs
passing the quality control criteria
were interrogated in an independent replication data set of
4218 cases and 7296 controls.

A methodologically challenging part of the study was the
non-UK component with 20,119 individuals of European ancestry
collected from 14 different countries.
Table~\ref{Table1} shows that
the case--control ratio varied strongly between the countries,
with some collections consisting only of case samples.
As a result, standard meta-analysis approaches,
where the samples from each country are analyzed separately and the
summary statistics combined, turned out to be inefficient.

%t1 #&#
\begin{table}
\caption{The origins of the samples in the non-UK component of the MS study}\label{Table1}
\begin{tabular*}{\textwidth}{@{\extracolsep{\fill}}lcclcc@{}}
\hline
\multicolumn{1}{@{}l}{\textbf{Country}}& \textbf{Cases} & \textbf{Controls} &\textbf{Country} & \textbf{Cases} & \textbf{Controls} \\
\hline
Finland & \phantom{0}581 & 2165 & Australia & \phantom{0}647 & -- \\
Sweden & \phantom{0}685 & 1928 & New Zealand & \phantom{0}146 & -- \\
Norway& \phantom{0}953 & \phantom{0}121 & Ireland & \phantom{00}61 & -- \\
Denmark & \phantom{0}332 & -- & USA & 1382 & 5370 \\
Germany & 1100 & 1699 & France & \phantom{0}479 & \phantom{0}347 \\
Poland & \phantom{00}58 & -- & Spain & \phantom{0}205 & -- \\
Belgium & \phantom{0}544 & --& Italy & \phantom{0}745 & \phantom{0}571 \\
\hline
\end{tabular*}
\end{table}

\begin{figure}

\includegraphics{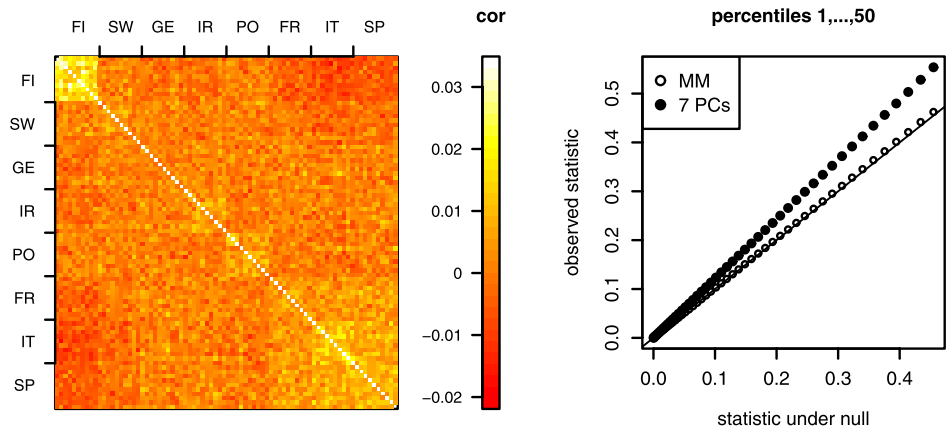}

\caption{{Left panel}. An $80\times80$ submatrix of the
genetic correlation matrix of
the non-UK individuals in the MS study. Ten randomly chosen individuals
are shown
for each of the following countries:
FInland, SWeden, GErmany, IReland, POland, FRance, ITaly and SPain.
The colors correspond to the pairwise correlation coefficients
according to the
scale in the middle. The diagonal values are close to 1.0 and are
colored white.
{Right panel.} The association test statistics of
470,000 SNPs
plotted from the 1st percentile to the 50th (median). The null distribution
on the $X$-axis is the chi-square with 1~df. Methods are
the linear mixed model (MM) and the logistic regression with 7 leading
principal components of the population structure as covariates (7PCs).
The line is $y=x$.}
\label{Fi}
\end{figure}

Alternative approaches, which jointly analyze data from several countries,
are likely to suffer from confounding effects of population structure.
Figure~\ref{Fi} shows a small part of a genome-wide correlation matrix
of the non-UK individuals calculated from about 200,000 SNPs.
Block-like structures on the diagonal show, unsurprisingly,
that the similarity of the genomes correlates with the
sampling locations.
Since the case--control status also has a strong dependence
on the sampling locations due to the ascertainment process (Table~\ref{Table1}),
spurious associations between
SNPs and the phenotype will arise
if the correlation structure in the data is not properly modeled.

%f1 #&#

We explored several approaches to address this issue.
First we conducted a meta-analysis on groups that had balanced
case--control ratios
and were genetically homogeneous, according to a model-based clustering
algorithm.
We also conducted logistic regression by including the seven leading
principal components
(PCs) of the population structure as covariates [\citet{Patterson2006}].
A standard way of checking GWAS analysis is based on the assumption that
only a very small proportion
of the variants affect the phenotype and, therefore,
the test statistics of the majority of the variants
should follow the null distribution [\citet{Devlin2001}].
This assumption is often assessed through the
``genomic control'' parameter, $\lambda$,
defined as the ratio of the median of the observed test statistic distribution
to that of the theoretical null distribution.
A substantial inflation was observed with
$\lambda=1.44$ for the clustering approach and
$\lambda=1.22$ for the PC approach (Figure~\ref{Fi}).
Although some of the inflation was likely to reflect the polygenic
architecture of the disease (small genetic effects at very many
variants) [\citet{Yang2011}],
it remained likely that the underlying population structure was
confounding the tests.

The linear mixed model as presented in this paper provided
a way to include the whole estimated
genetic correlation structure of 20,119 individuals in the regression model.
The model-checking confirmed that the confounding
effects were well controlled ($\lambda=1.02$, see Figure~\ref{Fi}) while
simultaneously
the method maintained power to detect associations,
as evidenced through the replication
of over 20 previously-known associations.
The main results of the MS GWAS, analyzed via the linear mixed model,
included the identification of 29 novel association signals.
These signals had important biological consequences, with
further analyses showing that
immunological genes are significantly overrepresented
near the identified loci.
In particular, the findings
highlight an important role for T-helper-cell differentiation
in the pathogenesis of MS.
Another striking pattern
was the very substantial overlap between genetic variants associated
with MS and those associated with autoimmune diseases
[see \citet{Sawcer2011} for further details].\looseness=1

%s3 #&#
\section{Binary data}\label{sec3}
The linear mixed model (\ref{eqgenerativemodel}) is formulated for
a univariate quantitative response
and, therefore, its application to binary case--control data requires
further justification.
A connection between the standard
linear model and the Armitage trend test [\citet{Armitage1955}] that we derive
in the supplementary text [\citet{Pirinen2012}]
adds to
the work of \citet{Astle2009} and \citet{Kang2010} who have previously used
the mixed model for significance testing in case--control GWAS.
In addition to testing, it is also important to measure the
effect sizes on a relevant scale.
Next we explain how the output from the standard linear model
can be turned into
accurate effect size estimates on the log-odds scale, which is
a natural scale for case--control studies.

For 0--1 valued responses $\bs Y=(y_1,\ldots,y_n)^T$ a
logistic regression model assumes that
%
%e3.1 #&#
\begin{equation}
p_i=P(y_i=1|\bs X,\bss\gamma)=\frac{\exp(\bs{X}_i \bss\gamma
)}{1+\exp(\bs{X}_i \bss\gamma)},
\end{equation}
where the row $i$ of $\bs X$ is denoted by $\bs X_i$
and the effects of the predictors are in the vector $\bss\gamma$.
The score function of the corresponding binomial likelihood
for a set of independent observations is
$\bs X^T(\bs Y - \bs p)$, where $\bs p=(p_1,\ldots,p_n)^T$ is a
function of $\bss\gamma.$
If we can justify a linear approximation
$\bs p \approx\bs X \bss\beta$, then
the score becomes approximately zero at the
least squares estimate
$\widehat{\bss\beta}=(\bs X^T \bs X)^{-1} \bs X^T \bs Y$.
In the supplementary text we argue that such an approximation is
good when the logistic model effects $\bss\gamma$ are small and we provide
a connection between the parameters $\bss\gamma$ and $\bss\beta$ in
those cases.
These steps allow us to use the output from the standard linear model
(i.e., the least squares solution $\widehat{\bss\beta}$) to
approximate the maximum likelihood estimates of the logistic regression model.
For our GWAS application,
where the case--control status is regressed on the population mean
and the (mean-centered) reference allele count at a SNP,
these considerations lead to
the following estimate of the genetic effect on the log-odds scale:
%
%e3.2 #&#
\begin{eqnarray}
\label{GWASappro} &&\widehat{\beta} \biggl(\phi(1-\phi
)+0.5(1-2\phi)
(1-2\theta)\widehat{\beta}
\nonumber
\\[-8pt]
\\[-8pt]
\nonumber
&&\qquad{}-\frac{0.084+0.9\phi(1-2\phi) \theta
(1-\theta)}{\phi(1-\phi)}\widehat{\beta}^2
\biggr)^{-1},
\end{eqnarray}
where $\phi$ is the proportion of the cases in the data, $\theta$ is the
reference allele frequency in the data and $\widehat{\beta}$ is the least
squares estimate of the effect of
the (mean-centered) reference allele count on the binary case--control status.

\begin{figure}

\includegraphics{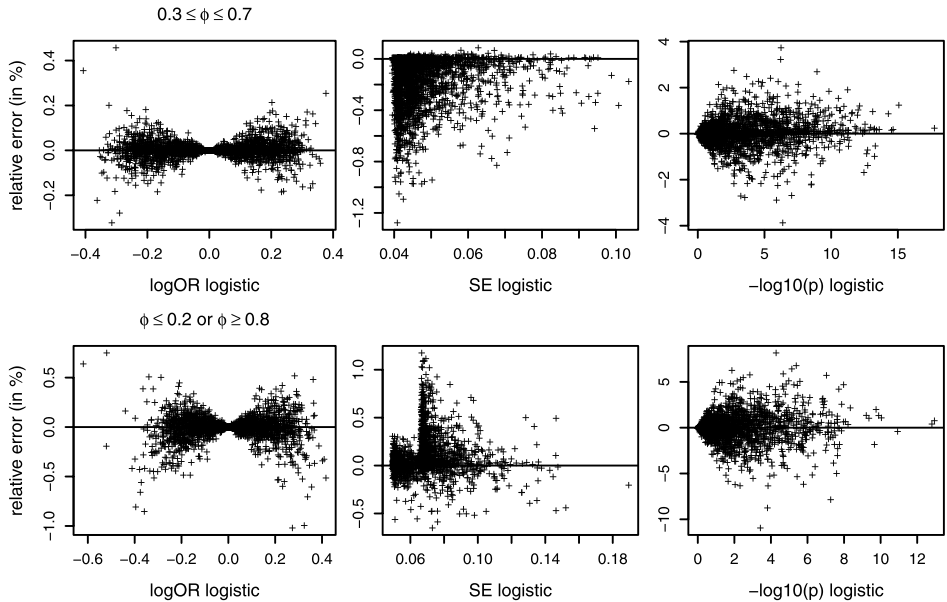}

\caption{Difference between the linear and the logistic model.
The panels include results from 2500 (top-row) and 2000 (bottom row)
binary variants simulated as described in the text.
The titles on the leftmost panels show the proportion of cases, $\phi
$, $y$-axes show the
relative differences between the linear and the logistic models in percentages
and $x$-axes show the results from the logistic model.
$\log \mathrm{OR}$, log-odds ratio; \textup{SE}, standard error; $-\log10(p)$, $-\log10$ of the
$p$-value from
the likelihood-ratio test.}
\label{Fig1}
\end{figure}

To investigate how well this approximation works in typical GWAS settings,
we simulated case--control data for 5000 unrelated individuals
at 500 SNPs
for nine case proportions $\phi\in\{0.1,0.2,\ldots,0.9\}$.
The allelic log-odds ratios $\gamma$ were taken from an equally spaced
grid on the interval
corresponding to odds ratios in $[1.0,1.3]$. This range
covers typical GWAS hits; for example, in our MS study the median
effect size among the 52 reported associations was 1.11 (minimum 1.08,
maximum 1.22).
In our MS study the lowest minor allele frequency among the variants
taken to replication
was 4.6\%, which motivated us to sample the risk allele frequencies for
the controls from
a $\operatorname{Beta}(2,2)$ distribution, truncated to the interval $(0.05,\ldots,0.95)$.
The frequencies in cases were determined by assuming that each copy of
the risk allele
increases log-odds of the disease additively by~$\gamma$.
Both linear and logistic regression models were then applied to the
data with
the population mean and the sampled genotypes as predictors.
The differences in log-odds estimates $\widehat{\gamma}$ and their
standard errors
together with the $p$-values
from the likelihood-ratio tests are shown in Figure~\ref{Fig1}, where
the parameter estimates from the linear model have been transformed
according to formula
(\ref{GWASappro}).

%f2 #&#

The conclusion from Figure~\ref{Fig1} is that
in a typical case--control GWAS data set where genetic effects are small,
the case--control ratio is well-balanced and allele frequencies are not extreme
(say, $\mathrm{OR} \leq1.3$, $0.30 \leq\phi\leq0.70$ and
$0.05<\mathrm{freq}<0.95$),
the standard linear model provides an accurate approximation of the
corresponding
logistic regression model. The relative errors in the
log-odds estimates or their standard errors
are at most around 1\% and in the $-\log 10$ $p$-values at most around 4\%
(top row of Figure~\ref{Fig1}).
This result is useful because it suggests a natural
way to apply the linear mixed model to binary data by using
generalized least squares estimates (details in the supplementary text).
The following empirical results show that
this procedure performed well in our application.

%f3 #&#
\begin{figure}

\includegraphics{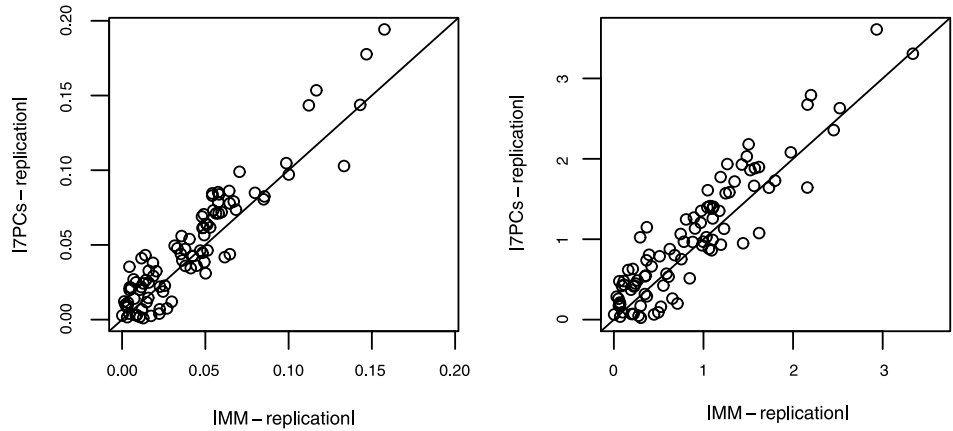}

\caption{Absolute differences of 93 effect sizes between
multiple sclerosis non-UK discovery and replication studies.
Scales are log-odds (Left panel) and
standardized log-odds (Right panel). MM: the linear mixed model in the
discovery data;
7PCs: logistic regression with
7 principal components of genetic structure as covariates in the
discovery data;
replication: the replication data
analyzed with logistic regression. Points above the diagonal:
$62/93$ (Left) and $59/93$ (Right).}\label{Fig2}
\end{figure}

In our multiple sclerosis study we took 93 independent SNPs
to the replication phase.
The replication analysis was conducted with 4218 cases and 7296 controls
using logistic regression [for details see \citet{Sawcer2011}].
Figure~\ref{Fig2} shows the absolute difference between the
effect sizes in replication analysis and in the non-UK part of the
discovery analysis using the linear mixed model ($x$-axes)
and logistic regression including 7 principal components
as covariates ($y$-axes). The log-odds ratios estimated by the linear
mixed model
were closer to the replication results
in 62 out of 93 SNPs (one-sided binomial $p$-value 0.0009).
The same pattern was present when the absolute differences are standardized
(59 out of 93, $p=0.006$), suggesting
that the methods presented here for estimating the
log-odds ratios by the linear mixed model can lead to more
accurate estimates than standard logistic regression
analyses when the data contain complex correlation structure which, for
practical
reasons, cannot be fully included in a logistic regression
model.

Obtaining effect size estimates and their standard errors is
critical in the genetics context both in interpreting the results
of individual studies and in combining results, via meta-analysis,
across studies.

%s4 #&#
\section{Maximum likelihood computation}\label{sec4}
The main analysis of our multiple sclerosis
study was based on maximum likelihood (ML).
In this section we consider how to efficiently maximize the likelihood function
corresponding to the sampling distribution
%
%e4.1 #&#
\begin{equation}
\label{eqlkhood} \bs Y | \bigl(\bss\beta, \sigma^2, \eta\bigr)
\sim
\mathcal{N}\bigl(\bs X \bss\beta, \eta\sigma^2 \bs R + (1-\eta)
\sigma^2 \bs I\bigr),
\end{equation}
with respect to $\bss\beta, \eta$ and $\sigma^2$.

In general, finding the ML estimates for linear mixed models
requires iterative procedures with expensive
matrix operations [\citet{Lynch1998}, page 787], but
for the particular model (\ref{eqlkhood})
more efficient algorithms can be found.
To our knowledge, the most efficient published algorithm is
EMMA [\citet{Kang2008}], which
has been applied to several recent GWAS [\citet{Atwell2010,Boyko2010}].
The algorithm FMM by \citet{Astle2009b},
which is currently being implemented
in the software suite GenABEL [\citet{Aulchenko2007b}],
was faster than EMMA in our test cases
but, to date, its exact computational details have not been published.
Another implementation of EMMA
is in the software package TASSEL (currently v.3.0) [\citet{Bradbury2007}],
which provides a graphical interface
and several approximations to reduce the running time.

Next we describe a novel conditional maximization algorithm which
is an order of magnitude faster than EMMA and was also faster than FMM
in our tests except with the smallest sample size of $n=250$ individuals.
We also consider in which situations
the full ML estimation is more powerful than
a recently proposed generalized least squares approximation [\citet
{Kang2010,Zhang2010}],
and compare the available methods.
Finally, we give running times on our MS data set.

%s4.1 #&#
\subsection{Conditional maximization}\label{sec4.1}
Our contribution to the ML estimation under the model (\ref{eqlkhood})
is a transformation of the data and predictors in such a way that
the covariance matrix becomes diagonal, enabling
an efficient conditional maximization procedure.
This transformation is a direct extension
of that used in general linear models to handle nondiagonal covariance matrices
to a more general case of two variance components.

The eigenvalue decomposition of the positive semi-definite matrix $\bs R$
yields an orthonormal $n\times n$-matrix
$\bs U$ of eigenvectors and a diagonal $n\times n$-matrix $\bs D$ of
nonnegative eigenvalues for
which $\bs R=\bs U \bs D \bs U^T$ [see \citet{Golub1996}].
Let us write $\til{\bs Y}=\bs U^T \bs Y, \til{\bs X} =\bs U^T\bs X,$ and
$\til{\bss\Sigma}= \eta\bs D + (1-\eta) \bs I.$
Then the log-likelihood function is
%
%e4.2 #&#
\begin{eqnarray}
\label{eqloglkhood} L\bigl(\bss\beta,\eta,\sigma^2\bigr)&=&c-
\frac{n}{2}\log\bigl(\sigma^2\bigr)-\frac{1}{2}\log\bigl(|\til{
\bss\Sigma}|\bigr)
\nonumber
\\[-8pt]
\\[-8pt]
\nonumber
&&{}-\frac{1}{2\sigma^2}(\til{\bs Y}-\til{\bs X} \bss
\beta)^T
\til{\bss\Sigma}^{-1} (\til{\bs Y} - \til{\bs X} \bss\beta),
\end{eqnarray}
where $c=-\frac{n}{2}\log(2\pi)$
and $|\til{\bss\Sigma}|$ denotes the determinant of $\til{\bss\Sigma}$
(details in the supplementary text).

Note that $\bs U,\bs D$ and $\til{\bs Y}$ are independent of $\bs X$
and that $\til{\bss\Sigma}$ is a
diagonal matrix which allows efficient computation of the inverse and
the determinant.
After the eigenvalue decomposition of matrix $\bs R$ [complexity is
$\mathcal{O}(n^3)$],
for each $\bs X$ the computation of $\til{\bs X}$ requires $\mathcal
{O}(k n^2)$ operations
where $k\leq K$ is the number of columns of $\bs X$ that
need to be recomputed,
and for each set of values of the parameters the evaluation of the
log-likelihood requires
$\mathcal{O}(nK)$ operations.
To maximize the log-likelihood, we apply a standard optimization
technique of
conditional maximization as described in the supplementary text.

%s4.2 #&#
\subsection{GLS approximation}\label{sec4.2}
In settings where the variance parameter $\eta$
does not vary much between the analyzed $\bs X$ matrices,
an efficient approximation can be found by
estimating $\eta$ only once and then applying a generalized least
squares (GLS)
method to approximate the ML estimates of $\bss\beta$ and $\sigma^2$
for any
given $\bs X$ matrix while $\eta$ is kept fixed.
This idea has been implemented in the software packages EMMAX [\citet{Kang2010}]
and TASSEL [\citet{Zhang2010}]; similar ideas had been proposed
earlier by \citet{Aulchenko2007}.
We will call this approach the GLS approximation to the full model.

The GLS approximation is accurate
only if $\eta$ does not vary much between different sets of predictors,
for example, when the individual predictors explain only a negligible
proportion of the total variance of the response.
This situation is typical in current GWAS studies on humans,
as the still unidentified genetic effects are small.
For example, in our MS study there were no noticeable differences between
the full likelihood analysis and the GLS approximation, and in their simulation
study \citet{Zhang2010} did not find significant differences in the
statistical power
between the two methods.
However, if the data contain closely related individuals
and individual genetic effects explain enough phenotypic variation, then
the full likelihood analysis may have higher power than the GLS
approximation, as we demonstrate below.
With our efficient implementation of the full model
it is possible to study this in more detail than before.

\subsubsection*{Family example}
We consider children of 25 independent families, each with~6 full-siblings
and a quantitative phenotype of whose variance 15\% is
explained by a major gene and 8.5\% by
minor genes (heritability is 23.5\%).
The remaining 76.5\% of the variation in the phenotype
is independent of the family structure.

We simulated 10 million such phenotypes and paired each with
a set of simulated genotypes that were independent of the phenotype
(given the family structure).
The minor allele frequency was chosen uniformly between 0.25 and 0.5,
and Hardy--Weinberg equilibrium [see, e.g., \citet{Lynch1998}] was assumed.
We used these data sets to get
accurate estimates of the threshold values of the
likelihood ratio statistic under the null hypothesis
of no genetic effect down to type I error $10^{-4}$.

We then simulated an additional one million phenotypes,
but this time tested the genotypes of the major gene
that influenced each phenotype. Using the empirical
threshold values (with their 95\% confidence intervals)
from the null simulations, the top two rows of Table~\ref{Table2} show the power of
the linear mixed model (MM),
the GLS approximation and the standard linear model (LM)
at type I errors $10^{-3}$ and $10^{-4}$.
In both cases MM is more
powerful than the GLS approximation, which
in turn is more powerful than LM.

%t2 #&#
\begin{table}
\caption{Power in family data. Columns: $\alpha$,
type I error rate; MM, linear mixed model; GLS, generalized least squares
approximation; LM, standard linear model. Cells give estimates of power
together with their 95\% confidence intervals. The first two rows
are based on the empirical type I error thresholds and the remaining
four rows use
theoretical thresholds}\label{Table2}
\begin{tabular*}{\textwidth}{@{\extracolsep{\fill}}lccc@{}}
\hline
\multicolumn{1}{@{}l}{$\bolds{\alpha}$}&\textbf{MM} & \textbf{GLS} &\textbf{LM} \\
\hline
\multicolumn{4}{c}{Empirical}\\
\multirow{2}{*}{$10^{-3}$}& 0.914 & 0.910 & 0.890 \\
&(0.913..0.915)&(0.909..0.911)&(0.889..0.891)\\
\multirow{2}{*}{$10^{-4}$}& 0.762 & 0.751 & 0.708 \\
&(0.757..0.767)&(0.746..0.757)&(0.702..0.714)\\ [3pt]
\multicolumn{4}{c}{Theoretical}\\
\multirow{2}{*}{$10^{-3}$}& 0.914 & 0.903 & 0.887 \\
&(0.913..0.914)&(0.903..0.904)&(0.886..0.887)\\
\multirow{2}{*}{$10^{-4}$}& 0.760 & 0.732 &0.702 \\
&(0.759..0.761)&(0.731..0.733)&(0.701..0.703)\\
\multirow{2}{*}{$10^{-6}$}&0.338 &0.293 &0.265 \\
&(0.337..0.339)&(0.292..0.294)&(0.264..0.266)\\
\multirow{2}{*}{$5\times10^{-8}$}& 0.145 & 0.115 &0.099 \\
&(0.144..0.145)&(0.114..0.115)&(0.098..0.099)\\
\hline
\end{tabular*}\vspace*{-3pt}
\end{table}

%t3 #&#
\begin{table}
\caption{Ratios of observed quantiles to expected. Columns: $\alpha$,
upper quantile; EXPECTED, theoretical 95\% confidence interval for the
ratio in $10^7$ samples; MM, linear mixed model; GLS, generalized least
squares approximation; LM, standard linear model. Values outside the
interval are in bold}\label{Table3}
\begin{tabular*}{\textwidth}{@{\extracolsep{\fill}}lcccc@{}}
\hline
\multicolumn{1}{@{}l}{$\bolds{\alpha}$}&\textbf{EXPECTED} & \textbf{MM} & \textbf{GLS} &\textbf{LM} \\
\hline
$10^{-3}$& 0.997..1.003 & 0.998 & \textbf{0.979} & \textbf{0.990} \\
$10^{-4}$& 0.992..1.008 & 0.997 & \textbf{0.973} & 0.992 \\
$10^{-5}$& 0.981..1.019 & 1.001 & \textbf{0.969} & 1.002 \\
\hline
\end{tabular*}
\end{table}

In practice, inferences in GWAS are based on the asymptotic
large-sample properties of the test statistics.
As mentioned in Section~\ref{sec2}, a
widely-used method for checking how\vadjust{\goodbreak} well the asymptotics hold
is to assess the ratio of the medians of the observed and expected
(chi-square) test statistic distributions, denoted by $\lambda$
[\citet{Devlin2001}].
For a sample of $10^7$ draws from the theoretical
null distribution the (analytically calculated)
upper bound of the 95\% confidence interval
of $\lambda$ is 1.0014, whereas in our $10^7$ null simulations we observed
values 1.055, 1.031 and 1.280 for MM, GLS approximation
and LM, respectively. Even though accounting for families
has brought MM and GLS much closer to the asymptotic
null distribution compared to LM,
both methods are still inflated with respect to the theoretical
distribution in this example
with a fairly small sample size.
Note that the GLS approximation always results in smaller
likelihood ratio statistics, and thus smaller $\lambda$ values,
than MM since GLS does not maximize the full model under the alternative
whereas MM does.

A simple way to make the observed test statistics match better with
the theoretical distribution is to divide them by
their corresponding estimates of $\lambda$,
a procedure called genomic control (GC) [\citet{Devlin2001}].
In this example GC works well but
since it treats all the variants the same,
it is not an ideal method for controlling for confounding
in more complex scenarios where different loci have
very different population genetic histories [\citet{Astle2009}].
Therefore, we have not used it with the MS data set.

Table~\ref{Table3} shows the ratios of some quantiles of the
observed distributions to their theoretical values
after genomic control, together with the theoretical
95\% intervals of those ratios assuming $10^7$ draws from the
null distribution. We observe no deviation from the theoretical
distribution for the linear mixed model and only a slight deflation for the
standard linear model, but the GLS approximation is deflated throughout
the range of quantiles considered.
Whether this phenomenon is specific to the family data considered or
holds more generally requires further investigation.
The lower panel of Table~\ref{Table2} shows power
at the theoretical thresholds corresponding
to type I error rates relevant in GWAS,
after genomic control was applied to the
one million nonnull tests.
The relative power difference between MM and the GLS approximation
increases with decreasing type I errors.

In this example MM was noticeably more powerful than the GLS approximation,
both at the empirical and theoretical thresholds, after
making the inflated statistics comparable by genomic control.
On the other hand, if neither empirical thresholds
nor genomic control parameters were available, then
the GLS approximation could be a more robust choice in small
data sets as reflected by
the observed~$\lambda$ values in this example.

%s4.3 #&#
\subsection{Comparing methods}\label{sec4.3}
Our conditional maximization (CM) algorithm, EMMA and the GLS
approximation all make use of
a decomposition of the $\bs R$ matrix requiring $\mathcal{O}(n^3)$ operations:
\begin{longlist}[(iii)]
\item[(i)] EMMA requires an additional $\mathcal{O}(n^3)$ matrix decomposition
for each set of predictors $\bs X$, whereas CM and GLS are $\mathcal{O}(n^2)$
algorithms for each $\bs X$ given the initial decomposition of $\bs R$.
\item[(ii)] EMMA reduces the problem to one-dimensional optimization for
which the global maximum
is in theory more reliably found than by using CM.
\item[(iii)] Parameterization of the model through $\eta$ (CM)
has a computational advantage over parameterization using $\delta
=(1-\eta)/\eta$
(EMMA) since the maximization is easier over the compact set $\eta\in[0,1]$
than over the unbounded interval $\delta\in[0,\infty)$.
\item[(iv)] It is expected that the GLS approximation is
computationally more efficient but in some cases
less accurate in ML estimation, and less powerful in testing
the predictors than either EMMA or CM, as demonstrated with the
previous family example.
\end{longlist}

We investigated through simulation studies
how the above differences
manifest themselves in practice, related to the reliability
and running time of the algorithms.
We applied the EMMA R-package v.1.1.2 [\citet{Kang2008}] with the
default parameters
and our C-implementation of the CM algorithm (software package MMM).
For the time comparisons we also included
a GLS approximation (our C-implementation in software package MMM) and
a beta version of the algorithm FMM\setcounter{footnote}{3}\footnote{Downloaded
in March 2011 from \url{http://astle.net/wja/}.}
[\citet{Astle2009b}].
We note that the software package TASSEL relies on the EMMA algorithm
in full ML estimation, and for the GLS approximation
both TASSEL and EMMAX are similar to our GLS implementation.
Therefore, TASSEL and EMMAX were not included in these comparisons.

%s4.3.1 #&#
\subsubsection{Reliability}\label{sec4.3.1}
The purpose of these tests is to assess whether condition~(ii) above
has any
practical effect on the variance parameter estimation.
For each value of $\eta\in\{0,0.05,0.1,\ldots,0.95,1\}$
we generated 1000 data sets for $n=500$ subjects.
A single data set consisted of an $\bs R$ matrix and a\vadjust{\goodbreak} $\bs Y$ vector.
To create $\bs R$, we simulated nonzero elements of an $n\times n$
lower triangular matrix $\bs L$ from the standard normal distribution and
set $\bs R= \bs L \bs L^T$ with the extra condition that if some of the
eigenvalues of
$\bs L \bs L^T$ were $<10^{-3}$ they were set to $10^{-3}$
to guarantee that~$\bs R$ was numerically positive-definite.
(The largest condition number of $\bs R$ matrices was $1.5\times10^{6}$.)
$\bs Y$ was then simulated according to the model
$\bs Y=\bss\varrho+ \bss\varepsilon$, where
$\bss\varrho\sim\mathcal{N}(0,\eta\bs R) $
and
$\bss\varepsilon\sim\mathcal{N}(0,(1-\eta) \bs I)$.
The ML estimates of $\eta$ and $\sigma^2$ were obtained from EMMA
and the CM algorithm. Since FMM does not output the
value of the maximized log-likelihood, we have not included it in this
comparison.
Also, for these data sets, the GLS approximation is the same as the full
model since we use each $\bs R$ matrix only once. Thus, no separate results
for GLS are reported.

%t4 #&#
\begin{table}
\caption{Maximum absolute differences between EMMA and CM and the
ranges of the
estimated quantities over 19,000 simulated data sets with $0.05 \leq
\eta\leq0.95$}\label{Table4}
\begin{tabular*}{\textwidth}{@{\extracolsep{\fill}}lccc@{}}
\hline
& \textbf{max log-likelihood} & $\bolds{\sigma^2}$ & $\bolds{\eta}$\\
\hline
$\max\Delta$ & 0.0053 & $3.0 \cdot10^{-5}$ &$ 1.2 \cdot10^{-5}$ \\
Range & ($-1618.332$, $-1085.829$) & (0.601, 1.444) & (0.029, 0.976) \\
\hline
\end{tabular*} \vspace*{-3pt}
\end{table}

The results for 19,000 data sets simulated with $0.05\leq\eta\leq0.95$
were the same between the methods for all practical purposes (Table~\ref{Table4}).
In addition to being similar up to 3 decimal places,
the optimized log-likelihood values
had no tendency of being higher with one algorithm than with the other
($p=0.46$ in the two-sided binomial test).

When $\eta$ was on the boundary $\{0,1\}$
the CM algorithm found points where the log-likelihood was at least
0.01 higher than that found by EMMA in 1503 cases out of the 2000
data sets
(maximum of these differences was 1.06).
This is due to property (iii) above, which requires EMMA to constrain
the search
to a compact subset of its unbounded search space. The size of the
search space is a
parameter of EMMA [we used the default values of $-10<\log(\delta
)<10$] and by increasing
this interval higher likelihood values could be found also by EMMA,
but with higher computational cost.
Alternatively, one could parameterize EMMA using $\eta$ instead of
$\delta$,
in which case EMMA and CM would be expected to give the same results
also on the boundary $\eta\in\{0,1\}$.

Thus, even if in theory the CM algorithm does not have guaranteed
convergence to the global optimum, in practice, it has found the same maxima
as EMMA in all 19,000 cases with $\eta\in\{0.05,\ldots,0.95\}$. Furthermore,
in the great majority of the remaining boundary cases $\eta\in\{0,1\}$
the CM algorithm has actually found a point with a higher
likelihood value than EMMA.
Since we generated the covariance matrices randomly without any
particular structure,
these results suggest that
the CM algorithm is a reliable method for the general problem of ML estimation
in the linear mixed model that we consider.\vadjust{\goodbreak}

%f4 #&#
\begin{figure}

\includegraphics{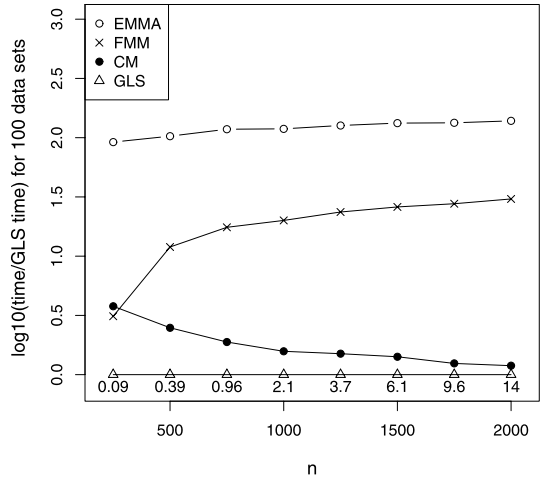}

\caption{Relative running times for 100 data sets compared to GLS, on the $\log 10$ scale,
as a function of the sample size $n$.
Methods are the R-package EMMA v.1.1.2,
our C-implementations of conditional maximization (CM)
and generalized least squares (GLS) and a C-implementation of FMM
(downloaded in March 2011).
The figures below the GLS-line are the GLS times in seconds.
Note that GLS is less accurate than the other three methods
which have fairly similar accuracy to each other.}\label{Fig4}
\end{figure}

%s4.3.2 #&#
\subsubsection{Running time}\label{sec4.3.2}
In applications, such as genome-wide association studies,
where a single covariance
matrix $\bs R$ is repetitively used with several sets of predictors
$\bs X$,
there is a large difference
in the running times between CM and EMMA due to property (i) above.
To investigate this difference, for each $n\in\{250,500,\ldots,2000\}$,
we simulated a single $\bs R$ matrix and $\bs Y$ vector as above,
together with 100 different $\bs X$ matrices.
Each $\bs X$ had dimension $n\times2$ and the
first column was always vector $\bs1$ to model the population mean
and the second column contained a randomly sampled binary vector where
each element was 1
with probability 0.5 and 0 otherwise.
The likelihood ratio (LR) tests for the effects $\beta_2$ were carried
out using
EMMA, FMM and our implementations of the CM and GLS algorithms.

Figure~\ref{Fig4} presents the running times as compared to
the GLS approximation.
We see that independently of the sample size, EMMA
takes about 100 times the time of the GLS procedure, reflecting the
fact that
EMMA carries out an additional
$n\times n$-matrix decomposition for each of the 100 data sets.

The relative efficiency of the GLS procedure over CM\vspace*{1pt}
decreases as the sample size grows, because both methods initialize
the data similarly by computing $\til{\bs X}=\bs U^T \bs X$, and
this task takes a larger and larger proportion of the whole running
time as $n$ grows.
A similar trend of decreasing relative difference is also present when the
initial matrix decomposition is subtracted from the running times of CM
and GLS
(results not shown).

The FMM algorithm is clearly faster than EMMA but slower than CM except
for the
smallest sample size $n=250$. We are not able to comment on the
putative sources of these
differences since the methodological details of FMM have
not been published.

Table~\ref{Table5} shows the maximum differences
between the methods in the likelihood ratio statistics and
the estimates of $\eta$.
We see that the results from EMMA and CM were again practically the same
over all 800 data sets and even though FMM deviated slightly from
the common results of EMMA and CM,
it was clearly closer to those two methods than to GLS.

%t5 #&#
\begin{table}
\caption{Maximum absolute pairwise differences between EMMA, CM, FMM
and GLS
in likelihood ratio statistic (upper diagonal) and $\eta$ estimate
(lower diagonal)
over the 800 data sets of Figure \protect\ref{Fig4}}\label{Table5}
\begin{tabular*}{\textwidth}{@{\extracolsep{\fill}}lcccc@{}}
\hline
&\textbf{EMMA} & \textbf{CM} & \textbf{FMM} & \textbf{GLS} \\
\hline
EMMA& -- &3.2$\times10^{-4}$& 0.089 & 0.43 \\
CM &8.1$\times10^{-6}$&-- &0.089 & 0.43 \\
FMM&0.0045\phantom{00.} &0.0046\phantom{000} &-- &0.34 \\
GLS &0.029\phantom{000.} &0.029\phantom{0000} &0.027 & --\\
\hline
\end{tabular*}
\end{table}

Given these results, it seems that CM is a natural choice
for likelihood inference in the linear mixed model (\ref{eqlkhood})
since it is much faster than EMMA,
more accurate than GLS and still computationally feasible
whenever GLS is.\looseness=-1

%s4.4 #&#
\subsection{MS data set}\label{sec4.4}
We applied the CM algorithm to the non-UK component of our
multiple sclerosis GWAS data set (20,119 individuals and 520,000 SNPs).
After the initial matrix decomposition was completed (in 3 hours 35 minutes),
the running time was 19 minutes 10 seconds per 1000 SNPs
using a single processor (Intel Xeon 2.50 GHz) and about 3GB of RAM, so
the whole MS data set can be run in 7 days and 2 hours
by using the CM algorithm on a single processor.
If instead one were to apply a method such as EMMA,
which requires a separate matrix decomposition at each SNP,
we estimate that the corresponding running time of
the whole MS data set would be about 210 years.
As noted earlier, in GWAS where genetic effects are small,
the GLS approximation (including programs EMMAX and TASSEL)
is expected to give, in practice, the same results as
the full likelihood analysis, and thus could also have been a possible
choice for this data set.
The running time of the GLS approximation on the
MS data set is about 5 days 19 hours, that is,
18\% less than that of CM.

%s5 #&#
\section{Bayes factors}\label{sec5}
A Bayesian framework provides a fully probabilistic quantification
of the association evidence, which is a useful complement to the
traditional frequentist interpretation in the GWAS context [\citet
{Wakefield2009}].
It also allows use of prior knowledge, for example,
a particular dependency between
the allele frequency and the effect size.
This possibility becomes more and more important as our
understanding about the genetic
architecture of complex traits develops [\citet{Stephens2009}].

In a Bayesian version of the linear mixed model (\ref
{eqgenerativemodel}), in addition
to the priors~(\ref{eqpriorsvar}) for the random effects, we adopt
the following priors:
\begin{eqnarray*}
\bigl(\bss\beta,\sigma^2\bigr) &\sim& \mathrm{Normal\mbox{-}Inverse\mbox{-}Gamma}(
\bs m,\bs V,a,b),
\\
\eta&\sim& \operatorname{Beta}(r,t).
\end{eqnarray*}
Here $a,b,r,t>0$ are scalar parameters, $\bs m$ is
a $K$-dimensional vector and
$\bs V$ is a $K\times K$ matrix.
In the supplementary text we describe the properties of these priors
and show how to efficiently evaluate the marginal likelihood of the data.
The marginal likelihoods allow comparisons between models that differ
in the structure of
the predictor matrix $\bs X$ (e.g., testing genetic effects in GWAS),
in the prior distributions of the parameters (e.g., whether $\eta=0$)
or both.

%s5.1 #&#
\subsection{Bayes factors for genetic association}\label{sec5.1}
\begin{figure}

\includegraphics{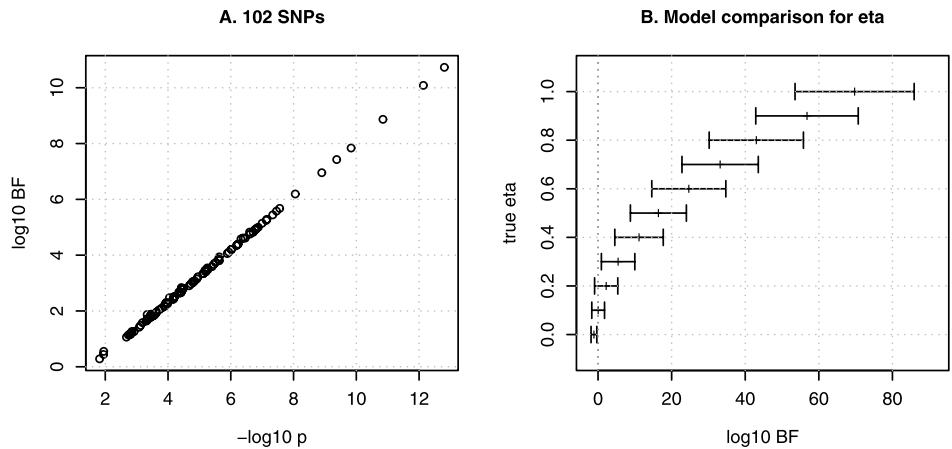}

\caption{\textup{(A)} Comparing $-\log 10$ $p$-values and $\log 10$ Bayes
factors in the non-UK
component of our MS study for 102 SNPs taken to replication.
\textup{(B)} Distribution of $\log 10$ Bayes factors between models
$\eta\sim\operatorname{Uniform}(0,1)$ and $\eta=0$.
For each value of $\eta\in\{0,0.1,\ldots,1\}$, 100 data vectors $\bs Y$
were simulated and means $\pm2\times$standard deviations of
the corresponding $\log 10 \mathrm{BF}$ distributions are shown. The proportions of
the data sets
for which $\log10(\mathrm{BF})>0$
were 0.01, 0.45 and 0.98, for true value of $\eta$ being
0.0, 0.1 and 0.2, respectively,
and 1 whenever $\eta\geq0.3$. }\label{Fig5}
\end{figure}

In the non-UK component of our MS data set
(20,119 individuals, 520,000 SNPs) the extra time spent in
computing the Bayes factors for SNP effects compared to computing
only the ML estimates
was 2 minutes 13 seconds per 1000 SNPs (Intel Xeon 2.50 GHz), that is,
an increase of about 12\% in the running time.
Following previous work [\citet{WTCCC2007}],
we chose the prior distribution on the
genetic effect to be centered at $0$ and have standard deviation
of 0.2 on the log-odds scale,
independently of the allele frequency. With this choice there was
nearly a linear relationship between the logarithmic $p$-values and
the logarithmic Bayes factors (Figure~\ref{Fig5}).
This data set-specific relationship provides useful information
about these two conceptually different quantities.

%f5 #&#

%s5.2 #&#
\subsection{Estimating heritability from a population sample}\label{sec5.2}
Recently, \citet{Yang2010} estimated the proportion of the variance
in human height that can be explained by
a dense genome-wide collection
of SNPs from a large sample of distantly related individuals.
Here we demonstrate Bayesian computation
by answering a related question of how high heritability
(i.e., $\eta$ in our mixed model)
needs to be in order to be detectably nonzero
from a particular sample of distantly related individuals.
Note, however, that we do not interpret $\eta$ as heritability
in our MS data set due to the confounding effects of the population structure.

We consider a sample of $n=5340$ UK individuals including
2665 healthy blood donors recruited from the United Kingdom Blood
Service (UKBS)
and 2675 samples from the 1958 Birth Cohort (1958BC).
These samples have been used as common controls for several GWAS
carried out by
the Wellcome Trust Case-Control Consortium 2.
Here we focus on the genotype data generated by the Affymetrix 6.0 chip.
After a quality control process, we made use of a genome-wide set of $S=168\mbox{,}351$
approximately independent SNPs
to compute a pairwise genetic correlation matrix $\bs R=(r_{ij})$ for
these individuals by setting
%
%e5.1 #&#
\begin{equation}
\label{relatedness} r_{ij}=\frac{1}{S}\sum
_{s=1}^S\frac{ (a_s^{(i)}-2p_s ) (a_s^{(j)}-2p_s )}{2p_s(1-p_s)},
\end{equation}
where $a_s^{(i)}$ is the number of copies of allele 1 that individual
$i$ carries at SNP $s$, $a_s^{(j)}$ is similarly defined for individual $j$,
and $p_s$ is the frequency of allele 1 at SNP $s$
in the whole sample of $n$ individuals.
The interpretation of $r_{ij}$ is that of relative genome-wide
sharing of alleles compared to an
average pair of individuals in the sample.
In particular, negative (positive) $r_{ij}$ denotes more distant (closer)
relatedness than that of an average pair in the sample, for whom the correlation
is $r_{ij}=0$.
The same matrix (divided by 2)
is called a ``kinship matrix'' by \citet{Astle2009}
and, excepting a slight adjustment on the diagonal,
it is also the same as the ``raw'' relatedness matrix used by \citet{Yang2010}.
For other versions of genetic relationship matrices, see, for example,
\citet{Kang2008,Astle2009}.
In our data all nondiagonal elements of $\bs R$ were below 0.03, showing
that there were no close relatives within this sample.

We simulated 100 phenotype vectors $\bs Y$ for the individuals
for each value of $\eta\in\{0,0.1,\ldots,1\}$ from the distribution
$\bs Y\sim\mathcal{N}(0,\eta\bs R+(1-\eta)\bs I)$.
We then compared two versions, $M_0$ and $M_1$, of the linear mixed model
\[
\bs Y=\beta+\bss\varrho+\bss\varepsilon\qquad \mbox{with }
\bss\varrho\sim\mathcal{N}(0,\eta\sigma^2\bs R) \mbox{ and }
\bss\varepsilon\sim\mathcal{N}\bigl(0,(1-\eta)\sigma^2\bs I\bigr),
\]
where in both models the prior on $(\beta,\sigma^2)$ was
$\operatorname{NIG}(m=0,V=10,a=10,d=12)$ and in $M_0\dvtx \eta=0$ and in
$M_1\dvtx \eta\sim\operatorname{Uniform}(0,1)$. For each data set
we computed marginal likelihoods $p(\bs Y | M_1)$ and $p(\bs Y | M_0)$
whose ratio gives the
Bayes factor (BF), which tells how the prior odds of the models are updated
to the posterior odds by the observed data $\bs Y$ [\citet{Kass1995}].
In particular, if BF${}>{}$1 [i.e., log10(BF)${}>{}$0], then the data favors
model $M_1$ over model $M_0$, and if BF${}<{}$1
[i.e., log10(BF)${}<{}$0], then the opposite is true. Figure~\ref{Fig5}
shows the distributions of log10(BF) for different (true) values of~$\eta$.
The running time of computing BFs for all 1100 data sets
was less than 5 minutes (Intel Xeon 2.50 GHz)
after $\bs R$ had been decomposed once, which took another 4 minutes.

We conclude that in our data set the model $M_0$ is (correctly) favored
in almost all the cases that were simulated with $\eta=0$ and that
when the true
$\eta\geq0.3$, then it is very likely that model $M_1$ will be
favored; that is,
in these individuals
we expect that this model comparison procedure detects nonzero heritability
for the phenotypes that truly have heritabilities $\eta\geq0.3.$
However, with real data there is a complication that the estimated
$\bs R$ matrix does not completely
capture the true genome-wide correlation, as only a subset of the
relevant variation
is used in estimating $\bs R$ [\citet{Yang2010}].
As a consequence, with real phenotype data
the lower limit of a convincingly detectable $\eta$ is likely to be higher
than in these simulations which have assumed that $\bs R$ is known exactly.

In general, the distribution of BFs depends on the sample size $n$,
the relatedness structure $\bs R$ and the priors on $\eta$,
and the formulae we have derived in the supplementary text
provide a computationally efficient way to
assess these dependencies in any particular data set.

%s6 #&#
\section{Discussion}\label{sec6}
Motivated by genome-wide association studies (GWAS),
we have presented computationally efficient ways to analyze
the linear mixed model
\begin{eqnarray}
\bs Y=\bs X \bss\beta+\bss\varrho+\bss\varepsilon\qquad
\nonumber\\
\eqntext{\mbox{with }\bss\varrho|\bigl(\eta, \sigma^2\bigr) \sim\mathcal{N}\bigl(0,
\eta\sigma^2 \bs R\bigr) \mbox{ and } \bss\varepsilon|\bigl
(\eta,
\sigma^2\bigr) \sim\mathcal{N}\bigl(0,(1-\eta)\sigma^2
\bs I\bigr)}
\end{eqnarray}
in situations where many
$\bs X$ matrices are analyzed with a single covariance matrix~$\bs R$.
In the GWAS context
the role of the random effect $\bss\varrho$ is to control
for those associations between phenotypes and genetic variants that can
already be explained
by the genome-wide genetic sharing.
The mixed model approach is especially useful when the\vadjust{\goodbreak} study individuals
show a complex relatedness structure which cannot be captured by including
a few linear predictors in the model.
Such a situation may arise if a case--control study combines individuals
from several populations with differing case--control ratios [e.g.,
\citet{Sawcer2011}]
or if the sampled individuals contain close relatives, for example, in
studies of
model organisms [\citet{Yu2005,Kang2008,Atwell2010}],
domesticated animals [\citet{Boyko2010}] or humans with
recent pedigree information [\citet{Aulchenko2007}].

For our case--control GWAS application [\citet{Sawcer2011}]
we have derived an accurate transformation between the
linear and logistic regression models when
the predictors have only small effects on the response.
This approach has the great benefit of enabling interpretation of the
linear mixed model results on
the log-odds scale, which is important in the GWAS context both for
understanding the sizes of the genetic effects and for combining the results
via meta-analyses across independent studies.

We have also formulated a conditional maximization (CM) algorithm for
maximum likelihood estimation
which is an order of magnitude faster
than the existing EMMA algorithm [\citet{Kang2008}]
and in our tests was also faster than the FMM algorithm [\citet{Astle2009b}],
except with the smallest sample size ($n=250$).
With the small effect sizes that are typical in current GWAS
the full mixed model analysis (performed by CM, EMMA and FMM)
gives very similar results to the generalized least squares
approximation (GLS)
that has been implemented in EMMAX [\citet{Kang2010}] and TASSEL [\citet
{Zhang2010}].
However, in other genetics contexts
the full mixed model may be more powerful than the GLS approximation,
as we
demonstrated with an example that
contained close relatives and genetic variants with large effects.
Given that our CM approach is computationally only slightly more
demanding than
the GLS approximation
(by about 20\% in running time in our large MS data set),
it seems well-suited for routine use in genetics applications.

We also considered computation of Bayes factors for
the genetic associations as well as for the variance components.
Another possible application of the Bayesian model
is in predicting an
unobserved response $y_i$ based on the set of observed values
$\bs Y_{-i}=(y_1,\ldots,y_{i-1},y_{i+1},\ldots,y_n)^T$
and the model $M$ (which contains information on priors, $\bs X$ and
$\bs R$).
The required posterior
$p(y_i|\bs Y_{-i},M)\propto p(y_i,\bs Y_{-i}|M)$ can be efficiently calculated
on a grid of possible values $y_i$ by using the methods described in
the supplementary text.
These calculations
are especially simple if response $y_i$ is restricted to
a set of discrete values as is the case
with binary data.

A natural question is whether the efficient computational solutions
presented in this article
could be extended to linear mixed models with more random effects,
as, for example, when analyzing gene expression data
with both the genetic relatedness and the expression heterogeneity
as random effects [\citet{Listgarten2010}].
The key issue that made the CM algorithm fast in our application
was the ability
to diagonalize the full covariance matrix $\bss\Sigma$ by using an
orthonormal matrix~$\bs U$
which did not depend on the variance parameters, or, in other words,
$\bs R$ and~$\bs I$ were simultaneously diagonalizable by the same
orthonormal $\bs U$.
More generally, a set of
symmetric matrices is simultaneously diagonalizable
by an orthonormal matrix
if and only if the matrices commute [\citet{Schott2005}, Theorem~4.18].
Thus, the computational strategy that we used here generalizes
straightforwardly only to
a rather special case of commutable covariance matrices.
In other situations an approximation to the full model could be
achieved by
the generalized least squares approximation where
the variance parameters are estimated only once and then
kept fixed for the repeated analysis of different predictor sets
[\citet{Kang2010,Zhang2010}].
On the other hand, an efficient generalization of both CM and EMMA to multiple
response vectors $\bs Y$ is straightforward since the necessary matrix
decompositions
do not depend on $\bs Y$.
This feature was utilized in our example of heritability estimation.

Extending linear mixed models to
proper variable selection models that simultaneously analyze
several thousands of predictors is also an important topic.
Further work is required to determine
whether the computational solutions presented in this work
can help implement more complex variable selection models.

Even though GWAS and other genetics applications
have given the main motivation for this study,
our results are more generally
valid for any application that fits into the framework of the
standard linear model with one additional normally distributed random effect.

\section*{Acknowledgments}
We thank the Area Editor and two referees for their helpful comments
that led to a considerable improvement of this paper.
We are also grateful to Dan Davison for his advice on
the matrix computations and to Davis McCarthy, C\'{e}line Bellenguez,
Gil McVean, Iain Mathieson and William Astle
for their comments on the manuscript.

\begin{supplement}[id=suppA]
\stitle{Supplementary text}
\slink[doi]{10.1214/12-AOAS586SUPP}  %[doi,text={...}] - jei reikia suskaldyti doi
\sdatatype{.pdf}
\sfilename{aoas586\_supp.pdf}
\sdescription{In this supplement we give the details of the application
of the mixed model to binary data,
of the conditional maximization of the likelihood function
and of the Bayesian computations.}
\end{supplement}

%
% imsref loaded by akundreckaite, 2012-09-06 13:20:30
% imsref loaded by akundreckaite, 2012-09-07 09:30:54

%suskaldyti doi

\printaddresses

\end{document}